\documentclass[aps,prb,twocolumn,showpacs,amsmath,amssymb]{revtex4-1}
\usepackage{graphicx}
\usepackage{dcolumn}
\usepackage{epstopdf}
\usepackage{bm}
\usepackage{bbold}

\begin{document}
\title{Correlation effects in insulating surface nanostructures}
\author{V.V. Mazurenko$^{1}$, S.N. Iskakov$^{1}$, A.N. Rudenko$^{1,2}$, I.V. Kashin$^{1}$, O.M. Sotnikov$^{1}$, M.V. Valentyuk$^{1,3}$, and A.I. Lichtenstein$^{3}$}
\affiliation{$^{1}$Theoretical Physics and Applied Mathematics Department, Ural Federal University, Mira Str.19,  620002
Ekaterinburg, Russia \\
$^{2}$ Institute of Chemical Reaction Engineering, Hamburg University of Technology, Eissendorfer Str. 38, 21073 Hamburg, Germany \\
$^{3}$Institute of Theoretical Physics, University of Hamburg, Jungiusstrasse 9, 20355 Hamburg, Germany}
\date{\today}

\begin{abstract}
 We study the role of static and dynamical Coulomb correlation effects on the electronic and magnetic properties of individual Mn, Fe and Co adatoms deposited on the CuN surface.  For these purposes, we construct a realistic Anderson model, solve it by using finite-temperature exact diagonalization method and compare the calculated  one-particle spectral functions with the LDA+$U$ densities of states. In contrast to Mn/CuN and Fe/CuN, the cobalt system tends to form the electronic excitations at the Fermi level.  Based on the calculated magnetic response functions, the relative relaxation times for the magnetic moments of impurity orbitals are estimated. To study the effect of the dynamical correlations on the exchange interaction in nanoclusters, we solve the two-impurity Anderson model for the Mn dimer on the CuN surface. It is found that the experimental exchange interaction can be well reproduced by employing $U$=3 eV, which is two times smaller than the value used in static mean-field LDA+$U$ calculations. This suggests on important role of  dynamical correlations in the interaction between adatoms on a surface. 
\end{abstract}

\maketitle

Recent scanning tunneling microscopy (STM) experiments on atoms deposited on surfaces have demonstrated many interesting and promising results varying from fundamental phenomena, such as the Kondo effect \cite{otte2} or giant magnetic anisotropy, \cite{gambardella} to practical realization of logic gates.\cite{Kha} Theoretical support of these experiments requires the construction and solution of realistic models that can capture the most essential couplings and effects in the system. In case of the impurity deposited on an insulating surface, such as Mn (Co, Fe) adatom on the CuN surface, it is natural to start the study with a spin-type Hamiltonian,\cite{fernandez,pruschke} where the impurity is considered as a localized spin.  This simple and attractive approach has been successfully applied to reproduce the Kondo effect in Co/CuN \cite{pruschke} or step-type excitations in the conductance spectra observed for Mn/CuN. \cite{fernandez, Fransson} 

The next step toward realistic modeling of Mn/CuN, Fe/CuN and Co/CuN systems would be to take into account the orbital nature of the adatom and hybridization effects between the adatom and surface states. This can be done by using a density functional theory (DFT) method \cite{bjones} or a many-body Anderson impurity model approach, or their combination. The first-principles DFT-based calculations provide an important information concerning ground state properties of the surface nanosystems. For instance, the densities of the states obtained  for Mn, Fe, Co/CuN systems by using the local density approximation (LDA), generalized gradient approximation (GGA) and local density approximation taking into account the on-site Coulomb interaction (LDA+$U$) demonstrated that the one-particle excitation picture depends on the symmetry of the particular $3d$ orbital of the adatom.\cite{bjones} 

On the other hand for the simulation of the electronic or magnetic excitations in a surface nanosystem, the Anderson model approach seems to be the most natural choice. The parameters for this model can be estimated by using the first-principles calculations. However, the solution of the Anderson model in case of the five-orbital impurity at experimental temperatures is still a challenging methodological and technical problem.

The correct description of the exchange interactions between impurity magnetic moments is another computational and methodological  problem. Numerous investigations based on the LDA+$U$ or GGA+$U$ approximations clarified the role of the static Coulomb interactions. The information about the influence of the dynamical Coulomb correlations on magnetic excitations in surface nanostructures is still limited.  For instance, the authors of Ref.\onlinecite{Kha2} proposed that a disagreement between theoretical and experimental estimates of the isotropic exchange interactions can be related to the lack of the dynamical correlation effects \cite{dynamical} that renormalize the electronic spectrum near the Fermi level. However, the proper microscopic mechanisms of correlated exchange process are still unknown.   

In this paper, we address above mentioned problems and study the correlation effects in transition metal atoms deposited on insulating CuN surface. We first analyze the spectral functions obtained from the LDA+$U$ and Anderson model calculations.
Based on the calculated spin-flip-type response functions of the impurity orbitals we discuss the origin of the low-energy excitations and estimate the relaxation times for magnetic moments of impurity orbitals. We clearly observe the difference in magnetic excitations between Co and Fe (Mn) systems. In contrast to the other, the cobalt system demonstrates an orbital polarization of the response functions.  To connect lifetimes of quantum spin-flip excitations with relaxation times of orbital magnetic moment, we used a combination of linear response theory and the Bloch equations approach. The effect of the dynamical correlations on the magnetic exchange interaction is demonstrated on the example of Mn dimer on the CuN surface.

{\it Methods.} In our study, we use two complementary approaches to investigate the electronic structure and magnetic properties of transition metal atoms deposited on the CuN surface: (i) the projected augmented-wave (PAW) method \cite{PAW} for individual Mn, Fe, or Co atoms, and (ii) the tight-binding linear-muffin-tin-orbital atomic sphere approximation (TB-LMTO-ASA) method \cite{OKA} for Mn dimer. The exchange and correlation effects have been taken into account by using the LDA and LDA+$U$\cite{Anisimov} approaches. The computational details are presented in the Supplementary materials. \cite{supplementary}

\begin{figure}[t]
\includegraphics[width=0.25\textwidth,angle=0]{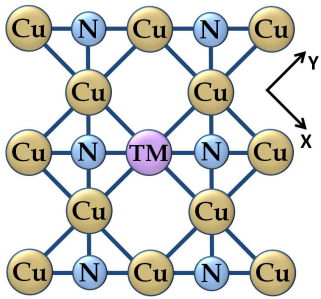}
\caption{ (Color online) Schematic representation of the individual transition metal atom (Mn, Fe and Co) on the CuN surface (top view) simulated in this work.}
\end{figure}

    \begin{table}[!tbp]
    \centering
    \caption[Bset]{Structural parameters for Mn, Fe, and Co adatoms supported on the CuN surface obtained from the LDA and LDA+$U$ ($U$=3, 6 eV)
calculations: the surface-adatom distance (in \AA) and the angle (in degrees) of nitrogen-metal-nitrogen bond. See text for details.}
    \label{str_param}
\begin{ruledtabular}
 \begin{tabular}{cccccc}
                               &     Mn             &     Fe     &    Co    \\
     \hline
  LDA                         &  1.49 (118)         &  1.46 (118)       &  1.43 (118)    \\
  LDA+$U$ ($U$=3 eV)   &  1.46 (123)         &  1.38 (129)       &  1.37 (125)     \\
  LDA+$U$ ($U$=6 eV)   &  1.53 (118)         &  1.43 (126)      &   1.40 (123)    \\
    \end{tabular}
\end{ruledtabular}
    \end{table}

{\it Geometry.}  The CuN(100) surface is built up from a (100) surface of fcc copper and nitrogen atoms uniformly embedded into the topmost layer
with the Cu:N ratio of 2:1 (Fig.1). The formation of covalent bonds in the Cu-N network favors an insulating behavior
of this surface.\cite{APL2007} The deposition of a single transition metal (TM) atom on top of a Cu atom results in a slight distortion
in the surrounding geometry due to formation of the N--TM--N bonds.\cite{rudenko} Equilibrium geometry for different single atom
adsorbates can be characterized by the distance from the adatom to the unperturbed topmost surface layer, and by the N--TM--N angle.
In Table \ref{str_param}, we list these quantities for Mn, Fe, and Co adatoms considered in this work. As one can see, in all three 
cases the geometries are similar, which suggests the same bonding mechanism for different adatoms. In contrast to the dimer and 
longer chains of TM atoms on CuN,\cite{rudenko} the on-site Coulomb correlations do not significantly affect the geometry of the 
system in the presence of a single atom on the surface.

\begin{figure}[h]
\includegraphics[width=0.4\textwidth,angle=0]{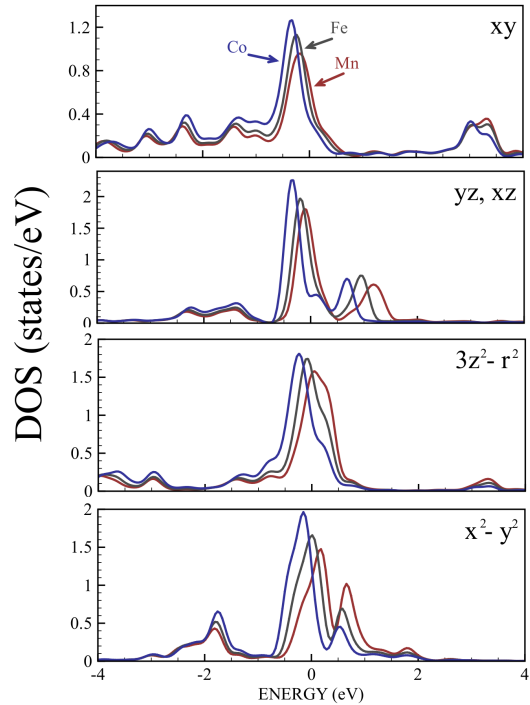}
\caption{ (Color online) Partial one-particle spectral functions of Mn/CuN (red), Fe/CuN (gray) and Co/CuN (blue) obtained from LDA calculations.}
\end{figure}

{\it LDA results.} First, we analyze spectral functions calculated by using the local density approximation (Fig.2). All the systems demonstrate similar one-particle excitation spectra with the shifted chemical potential due to the different number of 3d electrons.  The main feature of each spectrum is a peak at the Fermi level that originates from the hybridization between the adatom and surface states.  The width of the LDA peak of about 1 eV is approximately the same as for Co adatom on Pt(111), where strong hybridization and correlation effects were observed. \cite{Mazurenko} Thus the straightforward transition from the complex electronic Hamiltonian to a simple spin model is not justified.   

To give a quantitative estimate for LDA results and to extract the parameters for the many-body Anderson model, one can use the different schemes. One commonly used way is a  projection procedure \cite{proj} that defines the Wannier functions and the corresponding parameters for a tight-binding Hamiltonian. However, the energy bands of the surface nanosystems are strongly entangled in a wide range, which makes the band structure rather complicated. Moreover, since we use the exact diagonalization solver there is a limitation on the number of the surface orbitals. It is not clear which surface orbitals should be included in the Wannier function (WF) basis. Thus the construction of a moderate-size WF basis is not trivial for surface nanostructures. 

A more appropriate approach is to define some effective surface orbitals that reproduce main features of the LDA density of states.
It can be done within a minimization scheme where the non-interacting Green's function is diagonal in the orbital space and each LDA spectrum is fitted by the following expression 
\begin{eqnarray}
G^{LDA}_{i} (\omega)  = (\omega - \epsilon_{i} - \sum_{p = 1}^{2} \frac{|V_{ip}|^2}{\omega - \epsilon_{p}})^{-1},
\end{eqnarray}
where $\epsilon_{i}$ ($\epsilon_{p}$) is the energy of the impurity (effective surface) states and $V_{ip}$ is the hopping between i$th$ impurity and p$th$ effective surface orbitals. The calculated parameters are presented in Ref.\onlinecite{supplementary}. The typical values of the hybridization between the surface and impurity states for all the systems are ranging from 80 meV to 200 meV.

{\it The choice of $U$.}
The correct description of the localized nature of the 3d states of a transition metal adatom requires the proper account of the Coulomb correlation effects. It can be done by using the LDA+$U$ calculation scheme. In such calculations, one needs to specify the values of the on-site Coulomb and intra-atomic exchange  $J_{H}$.  There is a number of constrained procedures to calculate $U$ and $J_{H}$ within the DFT-based methods. However, depending on the approach that  can be local density approximation, generalized gradient approximation or random phase approximation (RPA), the resulting Coulomb interaction values are rather different. These values are also sensitive to the details of the surface. For instance, the constrained-LDA method gives U = 6.75 eV for individual Co atom on Pt(111) surface.\cite{Mazurenko} On the other hand the constrained-GGA results in U=0.8 eV in case of the Co/CuN system.\cite{bjones}

We also note that a common practice is to use the $U$ and $J_{H}$ as fitting parameters to achieve the best agreement with known experimental data. For instance, the LDA+$U$ method with $U$ = 5.88 eV gives the isotropic exchange interactions between Mn atoms on the CuN surface that agree well with experimental estimates.\cite{rudenko} As we will show below, this value of the U parameter can be substantially renormalized by taking into account the dynamical Coulomb correlation effects. Thus there is no unique set of the $U$ and $J_{H}$ parameters for a particular surface nanosystem.

In this work, we simulate the nanosurface systems in different interaction regimes with  two sets of the $U$ and $J_{H}$ parameters: $U$ = 3 eV ($J_{H}$ =0.9 eV) and $U$ = 6 eV ($J_{H}$ = 0.9 eV). The first set corresponds to nearly itinerant state with the effective Coulomb interaction ($U_{eff} = U - J_{H} =$ 2.1 eV), whereas the second can be associated with a localized regime ($U_{eff} = U - J_{H} =$ 5.1 eV). As the $U$ parameter of a single transition metal atom is known to decrease toward the surface, \cite{RudenkoX} two interacting regimes can be interpreted as two possible states, corresponding to different separations of the adatom from the surface. Moreover, comparing the Anderson model results for defined $U$ values will give us opportunity to retrace the corresponding changes of the electronic structure and magnetic properties.

{\it Anderson model results}
To simulate the many-body correlation effects in Fe, Co and Mn adatoms on the CuN surface, we use 
the single-impurity Anderson model that can be written in the following form:
\begin{eqnarray}
H =  \sum_{p \sigma} \epsilon_{p} c^{+}_{p \sigma} c_{p \sigma} 
+ \sum_{i \sigma}   (\epsilon_{i} - \mu) n_{i \sigma} \nonumber \\
+ \frac{1}{2} \sum_{i} g \mu_{B} B_{eff} (n_{i \uparrow} - n_{i \downarrow})  \nonumber \\
+  \sum_{i p \sigma} ( V_{i p} d^{+}_{i \sigma} c_{p \sigma} + H.c.) 
+ \frac{1}{2} \sum_{\substack {ijkl\\ \sigma \sigma'}}  U_{ijkl} d^{+}_{i \sigma} d^{+}_{j \sigma'} d_{l \sigma'} d_{k \sigma}. \, \,  \, \,
\end{eqnarray}
Here $\epsilon_{i}$ and $\epsilon_{p}$ are the energies of the impurity and bath states, $d^{+}_{i \sigma}$ and $c^{+}_{p \sigma}$ are the creation operators for impurity and surface electrons, respectively. $B_{eff}$ is an effective magnetic field applied in $z$-direction, $V_{ip}$ is the hopping between the impurity and surface states, $U_{ijkl}$ is the Coulomb matrix element and the impurity orbital index $i$ ($j$, $k$, $l$) runs over the $3d$ states. In our work, we use the full form of the Coulomb interaction matrix. For each system under consideration the Anderson model is solved by using the finite-temperature exact diagonalization method on the real energy axis.\cite{Mazurenko} 

It is important to discuss the magnetic field contributing to the Hamiltonian.  The experimental values of the magnetic fields (1-7 T) being used in our model can lead to nearly paramagnetic state with tiny magnetization \cite{Mazurenko2} or to magnetic solution with a low-spin state. It is mainly due to the use of the full Coulomb interaction matrix that contains spin-flip terms destroying the magnetic state. To obtain the magnetic solutions with the magnetic moment corresponding to the experimental spin, we need to use the effective magnetic fields of about 20-50 meV.  

\begin{figure}[b]
\includegraphics[width=0.5\textwidth,angle=0]{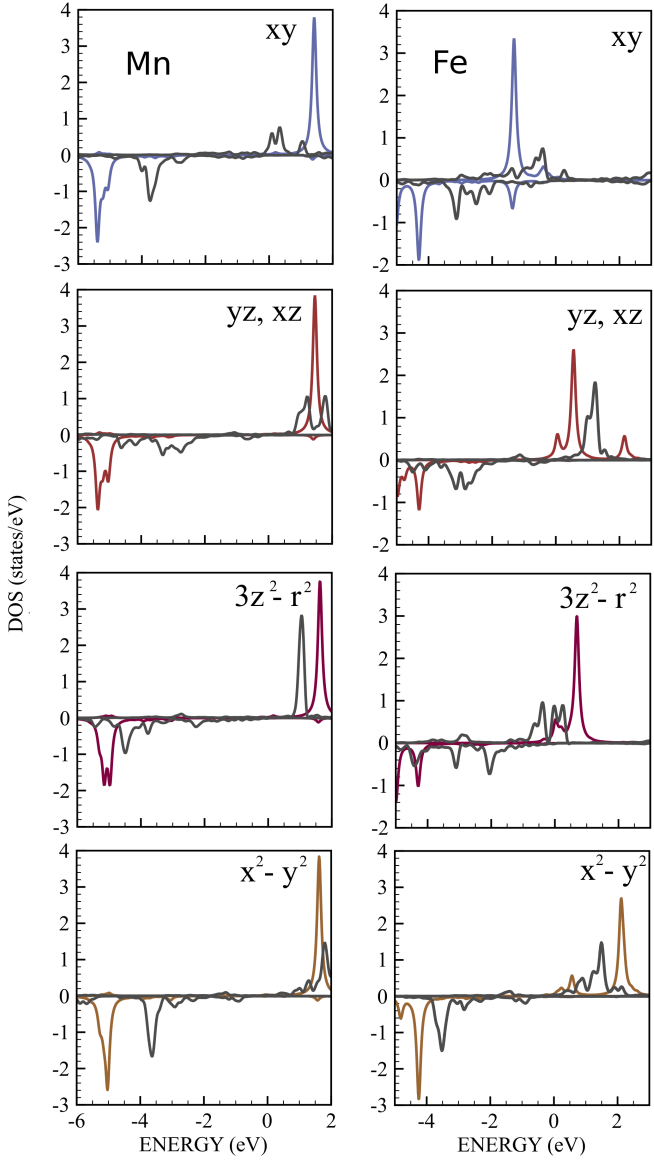}
\caption{ (Color online) Comparison of the partial spectral functions obtained from the LDA+$U$ (gray lines) and Anderson model calculations for the Mn/CuN (left) and Fe/CuN (right) systems ($U$=3 eV and $J_{H}$= 0.9 eV).}
\end{figure}

\begin{figure}[t]
\includegraphics[width=0.3\textwidth,angle=0]{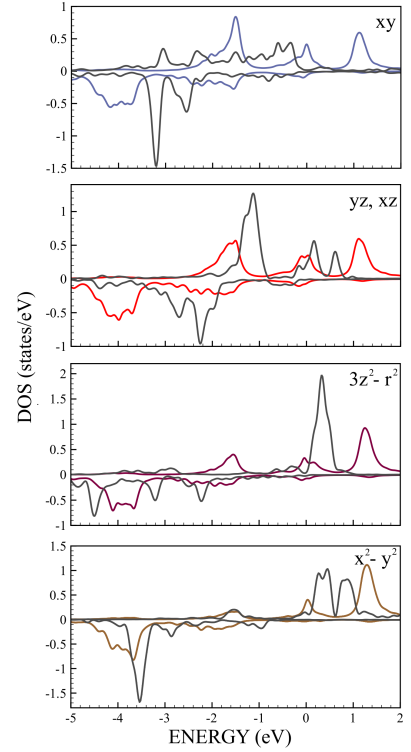}
\caption{ (Color online) Comparison of the partial spectral functions obtained from the LDA+$U$ (gray lines) and Anderson model calculations for the Co/CuN system ($U$=3 eV and $J_{H}$= 0.9 eV).}
\end{figure}

To assess the effect of the dynamical correlations, we compare the one-particle spectral function obtained from the Anderson model calculations with those calculated with the LDA+$U$ approach Fig.3 and 4. In case of Mn and Fe/CuN systems, we observe the tendency of the Anderson model solution to enhance the splitting between spin-up and spin-down states for partially filled orbitals. It is not the case for Co/CuN system, where the account of the dynamical correlations for Co adatom results in new excitations at the Fermi level.
In our model, the difference between Mn/CuN and Fe/CuN concerns the $xy$ orbital, which demonstrates a tiny magnetization in the case of Fe adatom. 
It is also important to note that we consider the magnetic solution of the Anderson model, which means that the correlation effects could be partially suppressed by magnetism. 

  \begin{table}[!tbp]
     \centering
     \caption[Bset]{Comparison of the magnetic moments (in $\mu_{B}$) of Mn, Fe, Co atoms obtained from the LDA+$U$ and Anderson model calculations with $U$ = 3 eV and $U$ = 6 eV.}
  \begin{tabular}{cccccc}
\hline
\hline
                       & Exp. & LDA+$U$     & LDA+$U$  & Anderson  & Anderson  \\
                       &          & $U$=3          & $U$=6        &  $U$=3      &  $U$=6           \\
      \hline
       Mn          & 5 ($S=\frac{5}{2}$) \cite{hirjibehedin}  & 4.0     &  4.4           &   4.5   &     4.7  \\
       Fe           & 4 ($S=2$) \cite{otte}                 &  3.0         &   3.2           &  3.7    &    3.7   \\
       Co          &3 ($S=\frac{3}{2}$) \cite{otte2}                  &  2.0         &   2.2           &   2.6   &   2.8         \\
\hline
\hline

     \end{tabular}
     \end{table}

{\it Spin-flip response functions}. To study the magnetic excitations, we have calculated the spin-flip response function for each impurity orbital using the Kubo's linear response theory,
\begin{eqnarray}
 \chi_{i}^{\mp} (\omega) = (g \mu_B)^2 \sum_{nn'}\frac{ |\langle n' | S^{+}_{i} | n\rangle|^2}{\omega - \Delta_{n'n} + i\Gamma (\Delta_{nn'})} (e^{-\beta E_{n}} - e^{-\beta E_{n'}}), 
 \end{eqnarray} 
 where $E_{n}$ and $| n \rangle$ are the eigenvalues and eigenstates of the Anderson Hamiltonian, and $\Delta_{n'n}= E_{n'} - E_{n}$ is an excitation energy. The corresponding spectral functions are antisymmetric,\cite{remark} i.e. they are positive for $\omega > 0$ and negative for $\omega < 0$. \cite{Kubo} Here we phenomenologically introduce the finite width for the collective spin excitations, $\Gamma (\Delta_{nn'})$ that relates to the lifetimes of these excitations, $\tau = \frac{\hbar}{\Gamma}$.  This allows us to decrease the effects of the discretized bath on the excitation spectrum. In general, the width can be calculated by using the self-energy part of the two-particle Green's function. To simplify the consideration, we use the principal property of $\Gamma$, that is  
 \begin{eqnarray}
 \Gamma (\Delta_{nn'}) << \Delta_{nn'}.
 \end{eqnarray}
The corresponding estimation for the ratio $\frac{\Gamma (\Delta_{nn'})}{ \Delta_{nn'}}$ can be found in the previous theoretical works. For instance, the authors of Ref.\onlinecite{Fransson} have fitted the experimental STM curve for Fe/CuN with the width $\Gamma$ = 0.01 - 0.03 meV. Taking into account the spin excitation energy of 1 meV, we can estimate  $\frac{\Gamma (\Delta_{nn'})}{ \Delta_{nn'}}$= 0.01. 

Since we simulate the magnetic states of the impurity with magnetic moments ranging from 2.2 eV (Co/CuN) to 4.4 eV (Mn/CuN), the intensity of the $zz$-component of the response function is strongly suppressed. Below we will analyze the $\chi^{\mp}_{i}$ and $\chi^{\pm}_{i}$ contributions that are related to each other by $Im(\chi^{\pm}_{i} (-\omega)) = -Im(\chi^{\mp}_{i} (\omega))$.  

\begin{figure}[b]
\includegraphics[width=0.5\textwidth,angle=0]{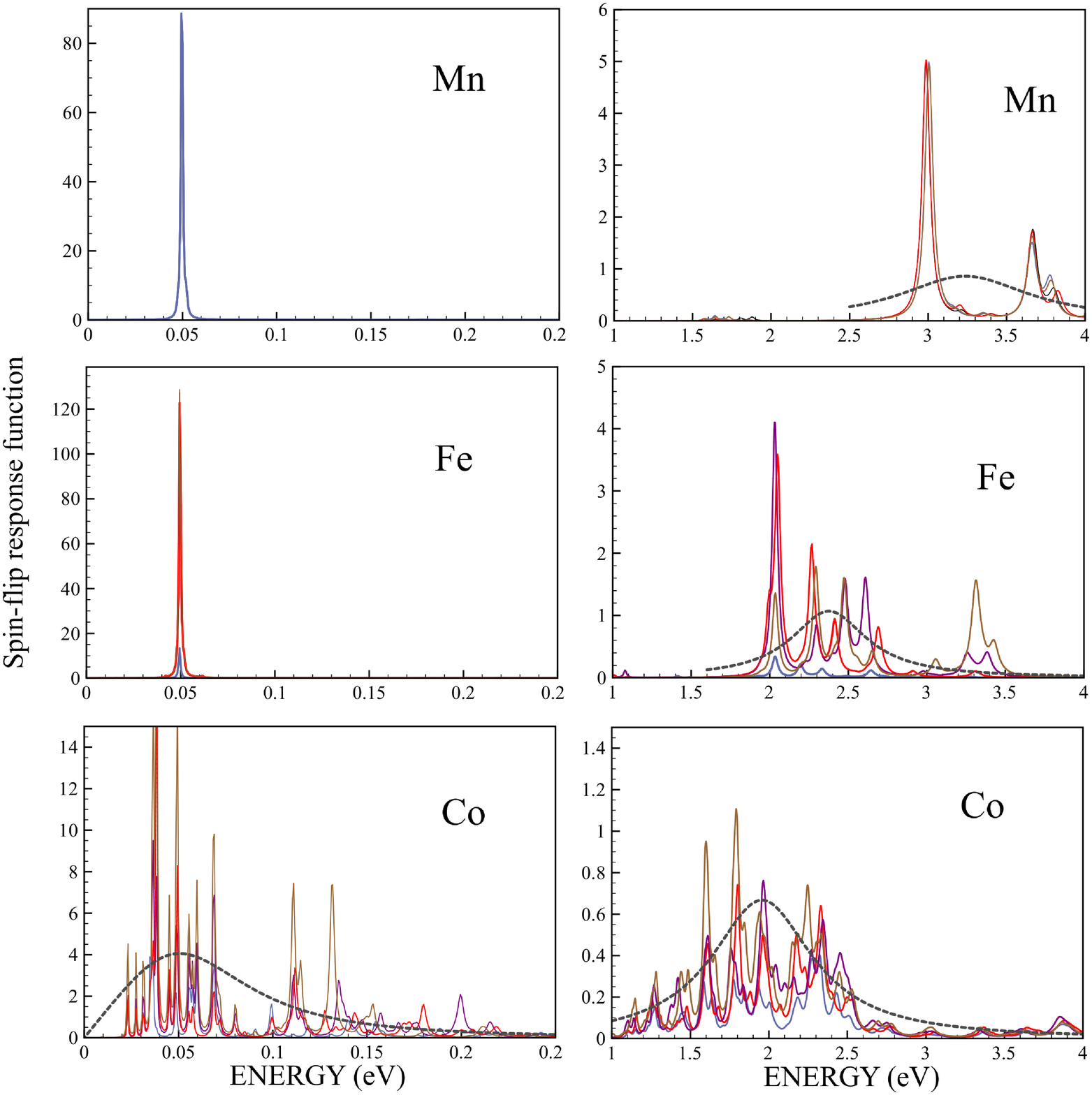}
\caption{ (Color online) Low- (left panel) and high-energy (right panel) parts of the spin-flip response functions (in $(g\mu_{B})^2 eV^{-1}$) calculated with $U$=3 eV and $J_{H}$= 0.9 eV. The blue, red, purple and brown lines correspond to the $xy$, $yz$ ($xz$), $3z^2-r^2$ and $x^2-y^2$ orbitals. The gray lines are examples of the fitting by using Bloch equation, Eq.(\ref{Bloch}).}
\end{figure}

The resulting spectral functions are presented in Fig.5. One can see that there are well-separated low- and high-energy excitations with the gap  ranging from 3 eV (Mn/CuN) to 1 eV (Co/CuN).  The low-energy peak is a response of the system on the external magnetic field, $B_{eff}$, as the resonance energy is fully controlled by the static $z$-component of the magnetic field, $B_{0}$. At the same time,  the high-energy structure corresponds to the intra-atomic spin-flip excitations. Analyzing these excitations at the time scale, it can be seen the intra-atomic excitations (1-2 eV) are typical for ultrafast processes at the femtosecond scale, whereas the response to the external magnetic field can be associated with the nanoscale range. 

Importantly, all the magnetic orbitals of Fe and Mn atoms demonstrate the same intensity of the low-energy peak. It is not the case for Co impurity where the intensity of the response function depends on the symmetry of the $3d$ orbital with the largest response resulting from the $x^2-y^2$ orbital. The smallest intensity of the low-energy peak is demonstrated by the $xy$ orbital. The difference between them is due to the hybridization with the nitrogen states. While the orbital of the $xy$ symmetry points in the direction of the largest overlap with nitrogen, the $x^2-y^2$ orbital lobes look away from nitrogen. 

{\it Orbital relaxation times}. For practical purposes it is important to show a connection between the lifetimes of quantum excitations and relaxation times of the magnetic moments of individual impurity orbitals.  For this purpose, we use the Bloch's \cite{bloch} approach and consider the nonequilibrium behavior of the orbital magnetic moments of the impurity in the magnetic field, $\vec B_{eff}$. This field has a stationary $z$-component of frequency $\omega_{0}$ = $g \mu_{B} B_{0}$ and weak time-dependent component along $x$ axis. Thus $\vec B_{eff}$ = ($B_{1}$ cos $\omega t$, 0, $B_{0}$). 

The impurity in the strong magnetic field applied along $z$ direction yields a finite magnetization. The $z$ component of the magnetization of the i$th$ orbital, $m^{z}_{i}$ fluctuates  near stationary solution $m^{0}_{i}$. Taking into account a small oscillating magnetic field in the $x$-direction  we obtain that the $x$ and $y$ components of the magnetization oscillate near the zero value.  If $m^{z}_{i}$ is perturbed from the equilibrium state ($m^{x}_{i}, m^{y}_{i} \ne$ 0) the return to this state will take characteristic times $T^{i}_{1}$ (longitudinal) and $T^{i}_{2}$ (transverse).
The corresponding equations of motion for the $x$ and $y$ components of the impurity orbital magnetization are given by
\begin{eqnarray}
\frac{dm^{x,y}_{i}}{dt} = [\vec m_{i} \times \vec B_{eff}]_{x,y} - \frac{m_{i}^{x,y}}{T^{i}_{2}}
\end{eqnarray}
The solution of this system of differential equations corresponds to the following susceptibility \cite{White}
\begin{eqnarray}
\tilde \chi^{xx}_{i} (\omega) = \frac{g \mu_{B} m^{0}_{i}}{2 T^{i}_{2}}[\frac{1}{(\omega^{i}_{0}-\omega)^2 + (1/T^{i}_{2})^2} - \frac{1}{(\omega^{i}_{0}+\omega)^2+ (1/T^{i}_{2})^2}] \label{Bloch}
\end{eqnarray}
where $m^{0}_{i}$ is the equilibrium value of the magnetic moment of the i$th$ orbital. Approximating nonequilibrium $\tilde  \chi^{xx}_{i} (\omega)$ by the  spin-flip susceptibilities obtained within the linear response theory, $\tilde  \chi^{xx}_{i} (\omega) = \frac{1}{4} (\chi^{\mp}_{i} + \chi^{\pm}_{i})$ we can estimate the orbital relaxation times. 

 \begin{table}[!tbp]
     \centering
     \caption[Bset]{Orbital relaxation times, $T^{i}_{2}$ estimated by minimizing Eq.(6).}
  \begin{tabular}{cccccc}
\hline
                       & $xy$  & $yz$     & $3z^2-r^2$  & $xz$  & $x^2-y^2$  \\
                       \hline
\multicolumn{6}{c}{Low-energy (slow) excitations}\\
      \hline
       Mn          & 7 ps  &  7 ps    &  7 ps          &   7 ps   &     7 ps \\
       Fe           & 6 ps    &  7 ps    &  7 ps          &  7 ps    &    7 ps   \\
       Co          & 40 fs  &  50 fs   &   60 fs           &    50 fs  &   80 fs        \\
\hline
\multicolumn{6}{c}{High-energy (fast) excitations}\\
\hline
       Mn          & 70 fs  &  70 fs    &  70 fs          &   70 fs   &   70 fs  \\
       Fe           & 17 fs  &  22 fs    &  15 fs          &  22 fs    &   11 fs   \\
       Co          & 9 fs  &  11 fs   &   9 fs           &  11 fs    &   12 fs        \\
\hline

     \end{tabular}
     \end{table}

The spin-flip response function contains low- and high-energy structures that correspond to different types of spin excitation in the system. We substitute these parts into Eq.(6) separately to simulate the magnetization dynamics on different times scales. In this case, the parameter $m^{0}_{i}$ controls the capacity of the corresponding susceptibility part and it can only be associated with the equilibrium magnetic moment of the impurity orbital. 
Let us start from the analysis of the low-energy excitations attributed to a response to the external magnetic field. The obtained relaxation times are presented in Table III. For Mn/CuN and Fe/CuN systems, we obtain the same relaxation times for the magnetic orbitals. In case of cobalt system, there is a small variation of $T_{2}$ for different impurity orbital states. Another important result is that the cobalt relaxation times are of two orders of magnitude faster than those of iron and manganese systems. It is due to the fact that the Co/CuN system is more itinerant than the others.

The spin-flip excitations of intra-atomic nature correspond to the high-energy part of the spectral function for $\omega >$ 1 eV.
The minimization procedure for this part of the spectrum leads to a picture of ultrafast relaxation times for the orbital magnetization (Table IV). The main difference from the low-energy excitation is that the Fe/CuN and Co/CuN systems demonstrate a complex cascade of the spin-flip excitations within the energy window of 1-2 eV. The manganese impurity shows two isolated excitations.
  
{\it Correlated exchange in Mn-dimer.} 
The next step of our study is to examine the influence of the dynamical correlations on the isotropic exchange interactions between adatoms on the surface. For these purposes we chose the Mn-dimer on the CuN surface \cite{hirjibehedin} that demonstrates the largest exchange interaction of 6.4 meV. The geometry of the Mn-dimer on the CuN surface is described in detail in our previous investigation.\cite{rudenko} 
To define the parameters of the two-impurity Anderson Hamiltonian, we use a combination of the projection procedure and minimization of the on-site LDA Green's function. The detailed discussion is presented in the Supplementary materials.  The constructed model is solved by using an exact diagonalization technique. The comparison of the calculated spectral functions with those obtained from the LDA+$U$ calculations is presented in Fig.7. One can see that, in contrast to LDA+$U$ results, the Anderson model demonstrates additional excitations at the Fermi level for the $3z^2-r^2$ and $xz$ orbitals.

\begin{figure}[t]
\includegraphics[width=0.3\textwidth,angle=0]{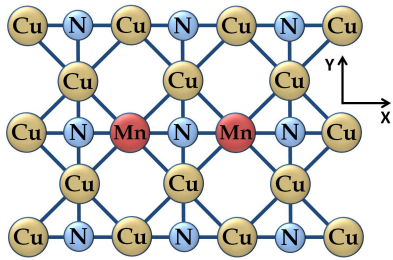}
\caption{ (Color online) Schematic representation of the Mn dimer on the CuN surface (top view) simulated in this work.}
\end{figure}

Before our discussing the two-impurity Anderson model and the results of its solution, let us analyze qualitatively possible effects of the dynamical Coulomb correlations on the exchange interaction. 
\begin{figure}[t]
\includegraphics[width=0.44\textwidth,angle=0]{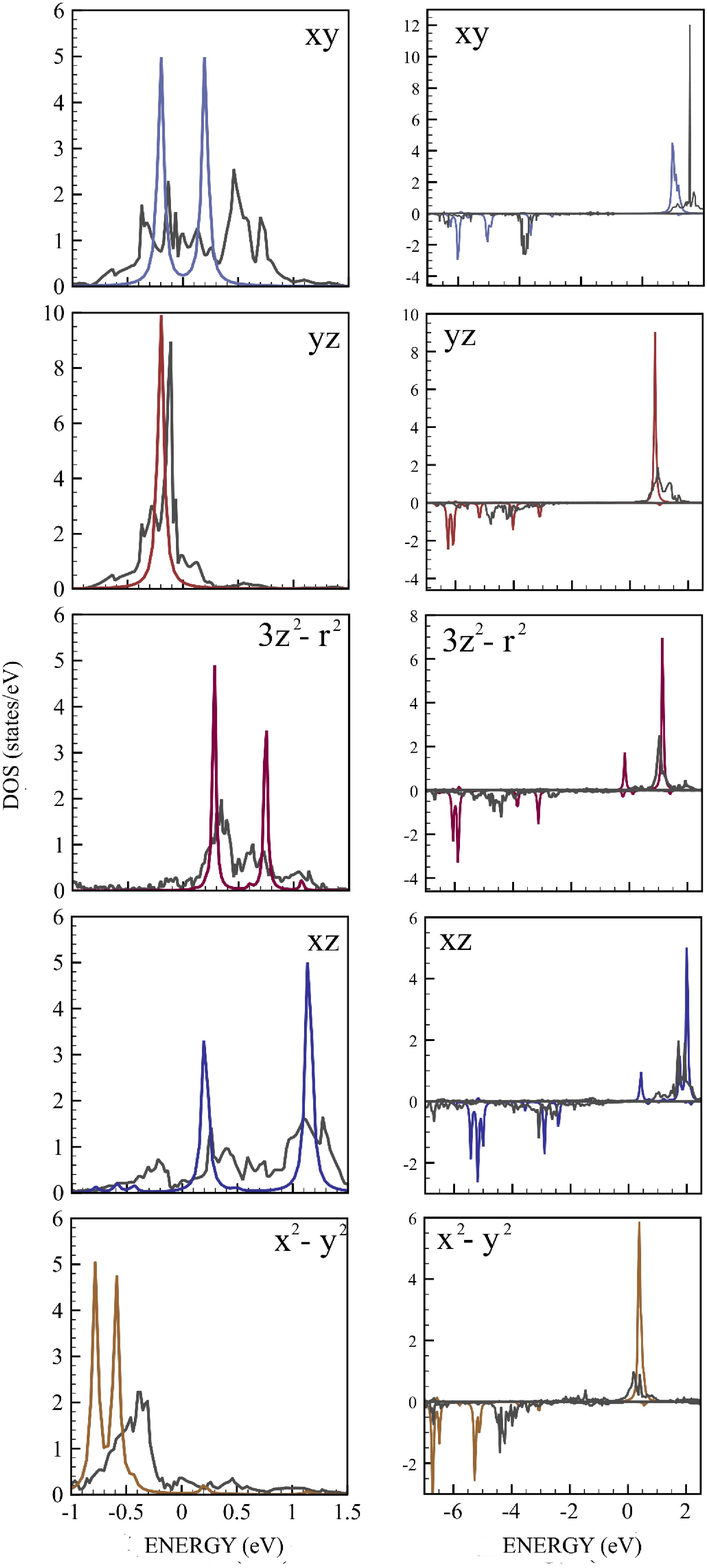}
\caption{ (Color online) Left panel. Comparison of the partial spectral functions obtained from LDA (gray lines) and non-interacting model Hamiltonian for the Mn dimer on CuN system. Right panel. Comparison of the partial spectral functions obtained from LDA+$U$ (gray lines) and Anderson model calculations for the Mn dimer on the CuN surface with $U$=3 eV and $J_{H}$= 0.9 eV.}
\end{figure}
For that we consider a general expression for the total exchange interaction of a given site with all magnetic environment,\cite{Liechtenstein, katsnelson, remarque}
\begin{eqnarray}
J_{0} = - \frac{1}{2 \pi S^2} Im \int_{-\infty}^{E_{F}} {\bf \Delta_0} (G_{\downarrow 0}  - G_{\uparrow 0} ) d \omega \nonumber \\
- \frac{1}{2 \pi S^2} Im \int_{-\infty}^{E_{F}}   {\bf \Delta_0} G_{\downarrow 0}  {\bf \Delta_0} G_{\uparrow 0} d \omega,
\label{exch LSDA}
\end{eqnarray}
where  $G_{\sigma 0}$ is the spin-dependent on-site local Green's function, the on-site potential ${\bf \Delta_0} = G^{-1}_{\downarrow 0}  - G^{-1}_{\uparrow 0}$ and $E_{F}$ is the Fermi energy. The Green's function taking into account the dynamical Coulomb correlations that can be obtained, for instance, in the LSDA+$\Sigma$ approach \cite{katsnelson} is written in the following form:
\begin{eqnarray}
\label{Green}
G_{\sigma 0}(\omega) = \sum_{{\bf k}} (\omega +\mu  - \epsilon^{\sigma} ({\bf k}) - \Sigma^{\sigma}_{0} (\omega) )^{-1},
\label{Green}
\end{eqnarray}
where $\sigma$ denotes spin, $\mu$ is the chemical potential, $\Sigma^{\sigma}_{0}$ is the self-energy and $\epsilon^{\sigma} ({\bf k})$ is the LSDA spectrum.
If the self-energy part in Eq.(\ref{Green}) is zero then we obtain the local LSDA Green's function. 

Let us compare the magnetic couplings calculated in LSDA (static mean-field) and LSDA+$\Sigma$ (dynamical mean-field) approaches.
For that, we expand the real part of the self-energy in the vicinity of the Fermi energy leaving only linear term, we obtain
\begin{eqnarray}
Re \Sigma (\omega) \approx \mu + (1- Z^{-1})  \omega,
\end{eqnarray}
where the quasiparticle residue is given by
\begin{eqnarray}
Z = (1- \frac{d Re \Sigma (\omega)}{ d \omega} |_{\omega = 0})^{-1}.
\end{eqnarray}
In turn, the correlated Green's function, Eq.(\ref{Green})  can be reduced to the effective LSDA form
\begin{eqnarray}
\label{G-m}
G_{\sigma 0} (\omega) \approx \sum_{{\bf k}}(\omega Z^{-1} - \epsilon^{\sigma}({\bf k}))^{-1}.
\end{eqnarray}
with rescaled frequencies.
This correlated Green's function being substituted into Eq.(\ref{exch LSDA}) leads us to the following ratio
\begin{eqnarray}
J^{LSDA+\Sigma}_{0} =  Z J^{LSDA}_{0}  
\label{stat-dyn}
\end{eqnarray} 
The performed qualitative consideration demonstrates that many-body correlation effects tend to renormalize a static mean-field (LSDA) value of the isotropic magnetic coupling. It is easy to show that the anisotropic interactions, such as Dzyaloshinskii-Moriya interaction and single-ion anisotropy are satisfied to the same relation (\ref{stat-dyn}) between static and dynamical values. Quantities related with the exchange interaction are also rescaled. For instance,  one can consider the Curie-Weiss  temperature that in the high-temperature limit can be written as $T_{CW} = \frac{J_{0}S(S+1) }{3 k_{B}}$.  

Since the impurity systems we study demonstrate a complex many-orbital behavior, it requires accurate numerical calculations. 
For reliable quantitative estimates of the magnetic coupling within the Anderson model approach, we used the exchange interaction in a strongly-correlated formulation, \cite{katsnelson}
\begin{eqnarray}
J_{12} =  \frac{1}{2 \pi S^2} Im \int_{-\infty}^{E_{F}} \Sigma^{s}_{1}  G^{\uparrow}_{12}  \Sigma^{s}_{2}  G^{\downarrow}_{21}  d \omega, \label{exchAnd}
\end{eqnarray}
where $\Sigma^{s}_{i} = \Sigma^{\uparrow}_{i}-\Sigma^{\downarrow}_{i}$ is the self-energy part of the on-site Green's function and $G^{\sigma}_{12}$ is the inter-site correlated Green's function.

We now turn to the analysis of the calculated exchange interaction between manganese impurities on CuN. In the previous work, \cite{rudenko} we demonstrated the effect of the static Coulomb correlations by comparing magnetic couplings calculated by employing LSDA and LDA+$U$ approaches. The account of the $U$ correction leads to a strong renormalization of the exchange interaction. Here one can see that the account of the dynamical correlations leads to an additional decrease of the exchange interaction between magnetic moments of the manganese atoms. We note that the approximately the same value of the total exchange interaction is obtained in the LDA+$U$ and Anderson model calculations with completely different values of the on-site Coulomb interaction, $U$=3 eV (Anderson) and $U$=5.88 eV (LDA+$U$).\cite{rudenko} 

\begin{figure}[!h]
\includegraphics[width=0.45\textwidth,angle=0]{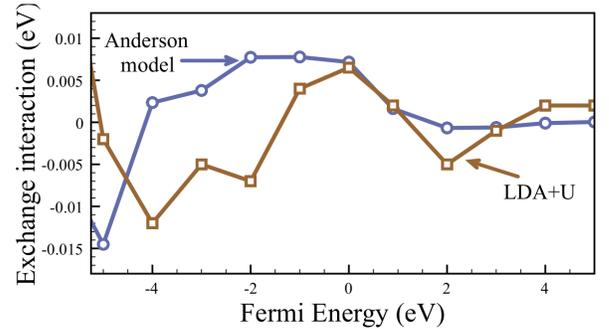}
\caption{ (Color online) Exchange interactions calculated in LDA+$U$ ($U$=5.88 eV) and Anderson model ($U$ = 3 eV)  as a function of the Fermi energy.}
\end{figure}

The decrease of the orbital contributions to the exchange interaction in LDA+$U$ and Anderson model calculations with respect to the LSDA results is not uniform. As can be seen from Table IV the interaction between the $xy$ orbitals of manganese atoms demonstrates the strongest renormalization. This allows us to conclude that the account of the frequency dependence of the self-energy plays a crucial role for the correct description of the magnetic couplings in surface nanostructures. The best agreement with the experimental value of 6.4 meV is achieved by using $U$=3.1 eV in the Anderson model calculations.

Since applying the voltage between the tip and sample in STM experiments one shifts the Fermi level, $E_{F}$ of the sample, it is of interest to examine the stability of the antiferromagnetic coupling with respect to $E_{F}$. Fig.8 gives the dependence of the exchange interaction on the upper integration limit in Eq.(\ref{exchAnd}). One can see that the deviation from the half-filled state in Mn-dimer leads to  ferromagnetic solutions both in the LDA+$U$ and Anderson model calculations.

  \begin{table}[!tbp]
     \centering
     \caption[Bset]{Total exchange interaction, $J_{12}$ and the largest orbital contributions (in meV) to the magnetic coupling obtained from the LSDA, LDA+$U$ ($U$=5.88 eV) \cite{rudenko} and Anderson model ($U$=3 eV) calculations.}
  \begin{tabular}{ccccc}
\hline
\hline
                            &  \, $d_{xy}${\bf -}$d_{xy}$     \,   &   \, $d_{3z^2-r^2}$ {\bf -} $d_{xz}$ \,      & \, $d_{xz}$ {\bf -} $d_{x^2-y^2}$  \,      &  $J_{12}$               \\
      \hline
       LSDA            &      5.9             & 1.7                                         &  2.7                                     &      15.3          \\
       LDA+$U$     &      1.8             &  1.2                                        &  1.7                                     &     6.5            \\
       Anderson  &      1.5             &  1.4                                      &  1.1                                    &   7.1         \\
\hline
\hline

     \end{tabular}
     \end{table}

To conclude, using the combination of the LDA and Anderson model calculations we have studied the electronic and magnetic excitations in Mn, Fe, and Co impurities deposited on the CuN surface. In the framework of the linear response theory, we observe fast intra-atomic spin-flip excitations at high energies and slow excitation that are due to the external magnetic field. The estimated relaxation times of the magnetic moments of 3d-Co orbitals are found to be two times smaller than that for iron and manganese adatoms on CuN. The two-impurity model calculations reveal a strong renormalization of of the exchange interaction between magnetic moments of Mn atoms  due to the dynamical correlation effects.  This result should be carefully taken into account for a realistic modeling of surface nanostructures. 

{\it Acknowledgements.}
We thank S. Loth, J. Wiebe, V.I. Anisimov and M.I. Katsnelson for helpful discussions.
The hospitality of the Institute of Theoretical Physics of Hamburg University is gratefully acknowledged.
This work is supported by DFG Grant  No. SFB 668-A3 (Germany), the grant program of President of Russian Federation
 MK-5565.2013.2, the scientific program ``Development of scientific potential of universities'', the contracts of the Ministry of education and science of Russian Federation N 14.A18.21.0076 and 14.A18.21.0889, Russian Foundation for Basic Research 12-02-31331, 12-02-31294, 13-02-00374.


\begin{thebibliography}{99}
\bibitem{dynamical}
By dynamical correlations we mean frequency dependence of the self-energy $\Sigma(\omega)$ (in contrast to the LDA+$U$ approximation where it represented by a static potential.)
\bibitem{remark}
Since the general force can not cause any effects prior to their action, our response functions correspond 
to the retarded two-particle Green's functions.
\bibitem{remarque}
The spin Hamiltonian is defined in the following form: $H = \sum_{i < j} J_{ij} \vec S_{i} \vec S_{j}$.
\end{thebibliography}
\end{document}